\documentclass[12pt]{article}
\usepackage[a4paper,textwidth=16cm,textheight=22.5cm,footnotesep=\baselineskip]{geometry}
\usepackage{bm}      
\usepackage{amsmath} 
\usepackage{cite}    
 \usepackage{slashed}
\usepackage{xcolor}
\usepackage{hyperref}
\hypersetup{final}
\hypersetup{
    colorlinks,
    linkcolor={red!50!black},
    citecolor={blue!50!black},
    urlcolor={blue!80!black}
                        }
\usepackage[utf8]{inputenc}
\usepackage{mathrsfs} 
\usepackage{calrsfs}
\usepackage{comment}

\newcommand{\gpri}{G}

\newcommand{\SU}{\text{SU}}
\newcommand{\tim}{\!\times \! } 
\newcommand{\tr}{{ \rm Tr} \,} 
\newcommand{\U}{\text{U}}

\begin{document}
\setlength{\baselineskip}{0.6cm} 


\newcommand*\xbar[1]{%
  \hbox{%
    \vbox{%
      \hrule height 0.5pt 
      \kern0.5ex
      \hbox{%
        \kern-0.1em
        \ensuremath{#1}%
        \kern-0.1em
      }%
    }%
  }%
}

\begin{centering}

  \textbf{\Large{
Remark on the QCD-electroweak phase transition         \\[2.5mm]
      in a 
supercooled Universe
}}

\vspace*{.6cm}

Dietrich B\"odeker
\footnote{bodeker@physik.uni-bielefeld.de%
} 

\vspace*{.6cm} 

{\em 
  Fakult\"at f\"ur Physik,
  Universit\"at Bielefeld,
  33501 Bielefeld,
  Germany
}

\vspace{10mm}

\end{centering}

\vspace{5mm}
\noindent

\begin{abstract}
In extensions of the Standard Model with no dimensionful parameters
the electroweak phase transition can be delayed to temperatures
of order 100 MeV. 
Then the chiral phase transition of QCD can proceed
with 6 massless quarks.
The top-quark condensate destabilizes the Higgs potential through the
Yukawa interaction, triggering the   electroweak transition.
Based on the symmetries of massless QCD,
it has been argued that the chiral phase transition is first order. 
We point out that the top-Higgs Yukawa interaction is 
nonperturbatively large at the QCD scale, and that its 
effect on the chiral phase transition may not be negligible,
violating some of the symmetries of massless QCD.
The remaining symmetries indicate that top quarks condense
in a second-order phase transition, but
top condensation might also be triggered by a first-order 
symmetry-breaking transition in the  
light-quark sector.

\end{abstract}


\section{Introduction}
First-order phase transitions are interesting for cosmology because
they can leave observable relics such as 
the baryon asymmetry of the Universe \cite{sakharov:1967dj,Kuzmin:1985mm}, 
gravitational 
waves \cite{Witten:1984rs}, or large-scale magnetic fields 
\cite{Vachaspati:1991nm}.
In the Standard Model there is neither an electroweak,
nor a QCD, or quark-hadron  phase transition: both electroweak symmetry 
\cite{Kajantie:1996mn,Csikor:1998eu}
and the approximate symmetries of
QCD get broken in a smooth crossover \cite{Aoki:2006we,HotQCD:2019xnw} 

However, in nearly scale-invariant extensions of the Standard Model
there can be a first-order electroweak phase transition, 
and, furthermore, it can have very peculiar features:
The decay rate of the metastable high-temperature phase
can be tiny down to very low temperatures, 
so that the Universe supercools
well below the critical temperature
\cite{Witten:1980ez,Espinosa:2007qk,Iso:2017uuu,Brdar:2019qut}. 
Important examples are Coleman-Weinberg type models \cite{Coleman:1973jx}, 
which contain no  
(negative) mass squared term for the Higgs field(s).
The thermal mass squared for the Higgs field $ H $ is positive, so that the 
state with $ H = 0 $ remains metastable until $ T = 0 $.
In some cases  the temperature becomes as low as $ O (   100 ) $ MeV while 
the electroweak symmetry is still unbroken, 
and all 6 quark flavors are still massless.
QCD with 6 massless quarks is expected to have
a first-order chiral phase transitions \cite{Pisarski:1983ms},
in which quark condensates 
$ \langle \overline q q \rangle \neq 0$ form 
at a critical temperature of about 85 MeV \cite{Braun:2006jd}.
If this also happens in the strongly supercooled Universe, 
the top condensate
would give rise to a linear term in the effective Higgs potential
through the top Yukawa interaction
\begin{align}
	{ \cal L } _ { t \mbox{-} \rm Yukawa }  = 
	 y _ t \overline Q _ {  3 } \widetilde H t _ R + { \rm H.c.} 
	,
	\label{tyuk}
\end{align}
where $ \overline Q _ 3 = ( \overline { t _ L } , \overline { b _ L}  ) $ 
is the doublet of third-family left-handed quarks, 
$ t _ R $ is the right-handed top quark, and 
$ \widetilde H = i \sigma  ^ 2 H ^ \ast $.
The linear term destabilizes the minimum of the effective potential
at $ H = 0 $ and triggers electroweak
symmetry breaking
\cite{Witten:1980ez}.
After the top condensation the Universe either 
evolves into a metastable 
state, followed  by another phase transition,  
or it could directly roll to the minimum of the effective potential.
Similarly patterns arise in other scale invariant extension,
such as Randall-Sundrum models%
\cite{Servant:2014bla,vonHarling:2017yew,Baratella:2018pxi}%
.

At a QCD phase transition the QCD coupling is large, 
and perturbation theory cannot be applied.
Nonperturbative lattice computations with light
quarks are very difficult (see \cite{ Lombardo:2014mda,Cuteri:2021ikv}
for simulations
with large number of flavors).
Insights into the nature of a phase transition 
can be gained from symmetry considerations. 
The Lagrangian of massless QCD with 6 quarks is invariant under chiral
$ \SU ( 6 ) _ L \tim \SU ( 6 ) _ R $ flavor transformations.  
A quark condensate breaks the axial subgroup, leaving a vector $ \SU ( 6 ) $
symmetry. 
If this happens in a second-order phase transition, 
then at the critical temperature some fields become effectively massless,%
\footnote{Or, equivalently, some correlations lengths diverge.}
and the corresponding effective theory  becomes
scale invariant.
Then the renormalization group for the couplings of this effective theory
has an infrared-stable fixed point.
The nonexistence of an infrared-stable fixed point, on the other hand,
is an indication that the phase transition is first order.
Near a second-order phase transition, 
the effective long-distance 
degrees of freedom are scalar fields describing the quark condensates.
The effective Lagrangian, and thus the beta functions are determined by the
symmetries.
The beta functions can be computed in the so-called $ \epsilon  $ expansion:
They are treated perturbatively near
$ d = 4 $ dimensions by writing $ d = 4 - \epsilon  $, 
and expanding in powers of $ \epsilon  $.
The effective theory describing the long-distance degrees of 
freedom near a second order is 3-dimensional, so the
$ \epsilon  $ expansion is applied to $ \epsilon  = 1 $. 
For massless QCD with
$ N _ f \ge 3 $ quark flavors one finds no infrared-stable fixed point,
which hints at a first-order chiral phase transition 
\cite{Pisarski:1983ms}. 
At nonzero, but small quark masses a first-order phase transition is 
expected, even though it has not been observed in 
lattice simulations so far 
\cite{deForcrand:2017cgb,Philipsen:2019rjq,Kuramashi:2020meg,Cuteri:2021ikv}. 

\section{Symmetries with top-Higgs Yukawa interaction}
Previous discussions of an electroweak phase transition triggered 
by top-quark condensation \cite{Iso:2017uuu}
were based on the symmetries of QCD alone.
However, neither Yukawa nor electroweak gauge interactions respect the
$ \SU ( 6 ) _ L \times \SU ( 6 ) _ R $ symmetry.
One may assume 
that electroweak interactions can be neglected, 
but the top Yukawa 
coupling 
becomes nonperturbatively large 
near the QCD scale
\cite{Flores:1983tk,Hambye:2018qjv}.
In Ref.~\cite{Hambye:2018qjv}
the running couplings from the  
3-loop renormalization group of the massless Standard Model are 
shown: the top Yukawa coupling grows as fast as the QCD coupling
and blows up when the renormalization scale approaches 
the QCD scale parameter $ \Lambda  _ { \rm QCD } $.%
\footnote{The Higgs self-coupling blows up as well,  
but it does not affect
the symmetries discussed here.}
Therefore, it is likely that the top Yukawa interaction in Eq.~(\ref{tyuk}) 
would affect the chiral phase transition.
When the Higgs field and the top Yukawa interaction
are included, the remaining symmetry is 
\begin{align} 
	\gpri	
	= \U ( 1 )_ B \! \times 	 \U( 4 )_  L  \! \times  \! \U ( 5 )_  R  
	\!  \times \! G _ t \! 
	\label{gpri} 
	\,\,,
\end{align}
where $  \U ( 1 ) _ B $ represents baryon number,
$ \U ( 4 ) _ L $ acts on $ u _ L, \ldots  , c _ L $, 
$ \U ( 5 ) _ R $ acts on $ u _ R, \ldots  , b _ R $, 
and 
$ G _ t = \!  \SU(2)\!\times \! \U(1) $ acts like the electroweak gauge
group, but only on $ Q _ 3 $, $ t _ R $, and $ H $.
\section{Phase transitions}
In a chiral phase transition $ \gpri $ is broken by expectation values of
\begin{align}
	\Phi  _ { ij } \sim \overline { q _ { L i } } q _ { R j } 
	\label{phiij} 
	.
\end{align}  
If such a phase transition was second order, 
some components of $ \Phi  $ 
would become effectively massless at the critical temperature.
Unlike in the $ \SU ( 6 ) _ L  \times \SU ( 6 ) _ R $ symmetric case
\cite{Pisarski:1983ms},
not all $ \Phi  _ { ij } $ become massless at the same temperature,
but only those  related by the symmetry $ \gpri $.
Thus the breaking of $ \gpri $ might 
proceed in a sequence of phase transitions.

When the top condensate forms, $ \Phi  _ { i 6 } $ with $ i=5,6 $ 
(corresponding to $ b $ and $ t$) 
gets a nonzero expectation value.
It transforms as a doublet under SU(2).%
\footnote{It can mix with
$  \widetilde H $, which has the same quantum numbers.}
The long-distance effective theory is equivalent to an O($ N $) model with
$ N =4 $,
which is known to have a second-order phase transition
(see e.g.\ \cite{ZinnJustin:1996cy}).
Thus, one would conclude that the breaking of $ G _ t $ through
top-quark condensation happens in a second-order phase transition.

However, this is not the only possibility. 
Coming from high temperatures, there could first be one or several
of the following phase transitions involving the 
light-quark sector, which break different symmetries: 
$ (  i ) $ when $  \Phi  _ { ij } $ with $ i=1,\ldots  ,4 $, $ j=1,\ldots  , 5 $
gets an expectation value,  or
$ ( ii ) $ $  \Phi  _ { ij } $ 
with $ i=5,6         $, $ j=1,\ldots  , 5 $,
or $ (iii)$ $  \Phi  _ { i6 } $ with $ i=1,\ldots , 4 $.
Under the symmetry-group $ G $ in Eq.~(\ref{gpri}),
these fields transform like  
$ \Phi  \to U ^ \dagger \Phi  V $ with $ U \in  \U(4 ) _ L $,
$ V \in  \U(5)_ R $ in case $ ( i ) $, 
with $ U \in  \SU(2 )\subset G _  t  $,
$ V \in  \U(5)_ R $ in case $ ( ii ) $.  
In case $ ( iii ) $ the field transforms like
$ \Phi  \to U ^ \dagger \Phi  $ with 
$ U \in \U ( 4 )_ L $. 
Then the long-distance behavior is again described by an
O$ ( N ) $ model, this time
with $ N =8 $, which has a second-order phase transition
\cite{ZinnJustin:1996cy}.
Results applicable to cases $ ( i ) $ and $ ( ii ) $
were obtained in 
Ref.~\cite{Pisarski:1980ix}, where  the fixed points for complex scalar fields 
$ \Phi  = ( \Phi  _ { ij } ) $, $ i = 1,\ldots  ,M $,
$ j = 1, \ldots  , N $,   with
the $ \U ( M ) \! \times \! \U ( N ) $-symmetric Lagrangian 
\begin{align}
	{ \cal L } = \tr ( \partial _ \mu  \Phi  ^ \dagger )
	( \partial ^ \mu  \Phi  ) 
	+ r \tr \Phi  ^ \dagger \Phi  
	+ \frac u4 \left ( \tr \Phi  ^ \dagger \Phi  \right ) ^ 2
	+ \frac v4 \tr \left ( \Phi  ^\dagger \Phi  \right ) ^ 2
	\label{L} 
\end{align}
were studied at leading order
in the $ \epsilon  $ expansion. 
There it was found 
that there is an infrared-stable fixed point for $  M N < 2 $, 
and that there are additional fixed points when 
$ ( N + M ) ^ 2  \ge 12 (M N -2 ) $,
i.e., for $ N > N _ +  $ or $ N < N _ -  $ 
with $ N _ \pm = 5 M \pm [ 24 ( M ^ 2 -1 ) ] ^{ 1/2 }$.
This indicates that there are 
no infrared-stable 
fixed points for $ ( M, N ) = (4 , 5 ) $ and
$ ( M, N ) = ( 2, 5 ) $, corresponding to case $ ( i ) $ and $ ( ii ) $,
respectively.
One would therefore conclude that in these two cases the phase transition is  
first order. 
The analysis of \cite{Pisarski:1980ix}
was extended to five \cite{Calabrese:2004uk}  
and recently to six loops \cite{Adzhemyan:2021sug}.
The $ \epsilon  $ expansion turns out to be not particularly 
well behaved for $ \epsilon  = 1 $.
The result of a resummation performed in  Ref.~\cite{Calabrese:2004uk} 
indicate that $  N _  + $ could be slightly less than 5  
when $ M = 2  $,
so  there might be an infrared-stable fixed point for $ ( M, N ) = ( 2,5 ) $.
This was confirmed in \cite{Adzhemyan:2021sug}.
On the other hand, 
Refs.~\cite{Pisarski:1980ix,Calabrese:2004uk,Adzhemyan:2021sug}  all
find that $ N _ + > 5 $ for $ M =4 $,  
indicating that there is a first-order phase transition for 
$ ( M, N ) = ( 4,5 ) $.

If there is a first-order phase transition as described
in the previous paragraph, 
there are two possibilities. 
The first is that in this phase transition  a top  
condensate emerges as well, 
not for symmetry reasons, but because there is a discontinuous change of 
the properties of the system.
That would mean that the top condensate forms in a first-order phase transition.
Another possibility is that the top condensate  forms  later 
in a second-order transition as described above.
Which of these two possibilities is realized can
probably not be inferred from symmetry considerations, 
but only from lattice simulations of the QCD + Higgs system.

If electroweak interactions cannot be neglected,
the $ \Phi  _ { ij } $ are not gauge invariant,
and cannot serve as an order parameter indicating the breaking of 
a global symmetry.
In particular, there is no {\it global} symmetry which gets broken 
when $ \bar t t $ gets an expectation value 
from which one could infer information about the existence or the order
of a  phase transition.%
\footnote{
This situation is similar to the breaking of the electroweak gauge symmetry
through the Higgs expectation value, and for the measured value of the Higgs
boson   there is no phase transition in the Standard Model.
}
The baryon number is violated due to electroweak sphalerons.
A remaining global symmetry is
$ \SU ( 2 ) _ Q \!\times \! \SU ( 2 ) _ u \! \times \! \SU ( 3 ) _ d $ 
acting on the left-handed quark doublets and the right-handed
up-type quarks of the first two families, and on all right-handed 
down-type quarks.
This symmetry can be broken by expectation values of 
the gauge invariant operators 
$
\overline Q _ \alpha  \widetilde H u _ j$ 
($ \alpha  =1,2 $;  $ j= 1,2 $), or 
$ 
\overline Q _ \alpha  H d _ j$ ($ \alpha  =1,2 $;
 $ j=1,2,3 $). 
In a phase transition in which these operators acquire  
an expectation value the Higgs field would also have to become nonzero. 
Then it would still have to overcome the barrier of the effective potential,
without assistance from the  top quark.  

\medskip 

\noindent {\bf  \large Acknowledgments } I would like to thank 
Frithjof Karsch,  Rob Pisarski, Laura Sagunski,
Wolfgang Unger, and Clemens Werthmann for useful discussions
or remarks.
This work was funded in part by
the Deutsche Forschungsgemeinschaft (DFG, German Research
    Foundation) – Project No.\ 315477589 – TRR~211.

 \appendix



\bibliographystyle{jhep}
\bibliography{references}

\providecommand{\href}[2]{#2}\begingroup\raggedright\begin{thebibliography}{10}

\bibitem{sakharov:1967dj}
A.~D. Sakharov, \emph{{Violation of CP Invariance, C asymmetry, and baryon
  asymmetry of the universe}},
  \href{http://dx.doi.org/10.1070/PU1991v034n05ABEH002497}{\emph{Pisma Zh.
  Eksp. Teor. Fiz.} {\bf 5} (1967) 32--35}.

\bibitem{Kuzmin:1985mm}
V.~A. Kuzmin, V.~A. Rubakov and M.~E. Shaposhnikov, \emph{{On the Anomalous
  Electroweak Baryon Number Nonconservation in the Early Universe}},
  \href{http://dx.doi.org/10.1016/0370-2693(85)91028-7}{\emph{Phys. Lett.} {\bf
  155B} (1985) 36}.

\bibitem{Witten:1984rs}
E.~Witten, \emph{{Cosmic Separation of Phases}},
  \href{http://dx.doi.org/10.1103/PhysRevD.30.272}{\emph{Phys. Rev.} {\bf D30}
  (1984) 272--285}.

\bibitem{Vachaspati:1991nm}
T.~Vachaspati, \emph{{Magnetic fields from cosmological phase transitions}},
  \href{http://dx.doi.org/10.1016/0370-2693(91)90051-Q}{\emph{Phys. Lett. B}
  {\bf 265} (1991) 258--261}.

\bibitem{Kajantie:1996mn}
K.~Kajantie, M.~Laine, K.~Rummukainen and M.~E. Shaposhnikov, \emph{{Is there a
  hot electroweak phase transition at m(H) larger or equal to m(W)?}},
  \href{http://dx.doi.org/10.1103/PhysRevLett.77.2887}{\emph{Phys. Rev. Lett.}
  {\bf 77} (1996) 2887--2890},
  [\href{https://arxiv.org/abs/hep-ph/9605288}{{\tt hep-ph/9605288}}].

\bibitem{Csikor:1998eu}
F.~Csikor, Z.~Fodor and J.~Heitger, \emph{{Endpoint of the hot electroweak
  phase transition}},
  \href{http://dx.doi.org/10.1103/PhysRevLett.82.21}{\emph{Phys. Rev. Lett.}
  {\bf 82} (1999) 21--24}, [\href{https://arxiv.org/abs/hep-ph/9809291}{{\tt
  hep-ph/9809291}}].

\bibitem{Aoki:2006we}
Y.~Aoki, G.~Endrodi, Z.~Fodor, S.~D. Katz and K.~K. Szabo, \emph{{The Order of
  the quantum chromodynamics transition predicted by the standard model of
  particle physics}},
  \href{http://dx.doi.org/10.1038/nature05120}{\emph{Nature} {\bf 443} (2006)
  675--678}, [\href{https://arxiv.org/abs/hep-lat/0611014}{{\tt
  hep-lat/0611014}}].

\bibitem{HotQCD:2019xnw}
{\scshape HotQCD} collaboration, H.~T. Ding et~al., \emph{{Chiral Phase
  Transition Temperature in ( 2+1 )-Flavor QCD}},
  \href{http://dx.doi.org/10.1103/PhysRevLett.123.062002}{\emph{Phys. Rev.
  Lett.} {\bf 123} (2019) 062002},
  [\href{https://arxiv.org/abs/1903.04801}{{\tt 1903.04801}}].

\bibitem{Witten:1980ez}
E.~Witten, \emph{{Cosmological Consequences of a Light Higgs Boson}},
  \href{http://dx.doi.org/10.1016/0550-3213(81)90182-6}{\emph{Nucl. Phys.} {\bf
  B177} (1981) 477--488}.

\bibitem{Espinosa:2007qk}
J.~R. Espinosa and M.~Quiros, \emph{{Novel Effects in Electroweak Breaking from
  a Hidden Sector}},
  \href{http://dx.doi.org/10.1103/PhysRevD.76.076004}{\emph{Phys. Rev. D} {\bf
  76} (2007) 076004}, [\href{https://arxiv.org/abs/hep-ph/0701145}{{\tt
  hep-ph/0701145}}].

\bibitem{Iso:2017uuu}
S.~Iso, P.~D. Serpico and K.~Shimada, \emph{{QCD-Electroweak First-Order Phase
  Transition in a Supercooled Universe}},
  \href{http://dx.doi.org/10.1103/PhysRevLett.119.141301}{\emph{Phys. Rev.
  Lett.} {\bf 119} (2017) 141301},
  [\href{https://arxiv.org/abs/1704.04955}{{\tt 1704.04955}}].

\bibitem{Brdar:2019qut}
V.~Brdar, A.~J. Helmboldt and M.~Lindner, \emph{{Strong Supercooling as a
  Consequence of Renormalization Group Consistency}},
  \href{http://dx.doi.org/10.1007/JHEP12(2019)158}{\emph{JHEP} {\bf 12} (2019)
  158}, [\href{https://arxiv.org/abs/1910.13460}{{\tt 1910.13460}}].

\bibitem{Coleman:1973jx}
S.~R. Coleman and E.~J. Weinberg, \emph{{Radiative Corrections as the Origin of
  Spontaneous Symmetry Breaking}},
  \href{http://dx.doi.org/10.1103/PhysRevD.7.1888}{\emph{Phys. Rev.} {\bf D7}
  (1973) 1888--1910}.

\bibitem{Pisarski:1983ms}
R.~D. Pisarski and F.~Wilczek, \emph{{Remarks on the Chiral Phase Transition in
  Chromodynamics}},
  \href{http://dx.doi.org/10.1103/PhysRevD.29.338}{\emph{Phys. Rev.} {\bf D29}
  (1984) 338--341}.

\bibitem{Braun:2006jd}
J.~Braun and H.~Gies, \emph{{Chiral phase boundary of QCD at finite
  temperature}},
  \href{http://dx.doi.org/10.1088/1126-6708/2006/06/024}{\emph{JHEP} {\bf 06}
  (2006) 024}, [\href{https://arxiv.org/abs/hep-ph/0602226}{{\tt
  hep-ph/0602226}}].

\bibitem{Servant:2014bla}
G.~Servant, \emph{{Baryogenesis from Strong $CP$ Violation and the QCD Axion}},
  \href{http://dx.doi.org/10.1103/PhysRevLett.113.171803}{\emph{Phys. Rev.
  Lett.} {\bf 113} (2014) 171803}, [\href{https://arxiv.org/abs/1407.0030}{{\tt
  1407.0030}}].

\bibitem{vonHarling:2017yew}
B.~von Harling and G.~Servant, \emph{{QCD-induced Electroweak Phase
  Transition}}, \href{http://dx.doi.org/10.1007/JHEP01(2018)159}{\emph{JHEP}
  {\bf 01} (2018) 159}, [\href{https://arxiv.org/abs/1711.11554}{{\tt
  1711.11554}}].

\bibitem{Baratella:2018pxi}
P.~Baratella, A.~Pomarol and F.~Rompineve, \emph{{The Supercooled Universe}},
  \href{http://dx.doi.org/10.1007/JHEP03(2019)100}{\emph{JHEP} {\bf 03} (2019)
  100}, [\href{https://arxiv.org/abs/1812.06996}{{\tt 1812.06996}}].

\bibitem{Lombardo:2014mda}
M.~P. Lombardo, K.~Miura, T.~J. Nunes~da Silva and E.~Pallante, \emph{{One,
  two, zero: Scales of strong interactions}},
  \href{http://dx.doi.org/10.1142/S0217751X14450079}{\emph{Int. J. Mod. Phys.
  A} {\bf 29} (2014) 1445007}, [\href{https://arxiv.org/abs/1410.2036}{{\tt
  1410.2036}}].

\bibitem{Cuteri:2021ikv}
F.~Cuteri, O.~Philipsen and A.~Sciarra, \emph{{On the order of the QCD chiral
  phase transition for different numbers of quark flavours}},
  \href{https://arxiv.org/abs/2107.12739}{{\tt 2107.12739}}.

\bibitem{deForcrand:2017cgb}
P.~de~Forcrand and M.~D'Elia, \emph{{Continuum limit and universality of the
  Columbia plot}}, \href{http://dx.doi.org/10.22323/1.256.0081}{\emph{PoS} {\bf
  LATTICE2016} (2017) 081}, [\href{https://arxiv.org/abs/1702.00330}{{\tt
  1702.00330}}].

\bibitem{Philipsen:2019rjq}
O.~Philipsen, \emph{{Constraining the phase diagram of QCD at finite
  temperature and density}},
  \href{http://dx.doi.org/10.22323/1.363.0273}{\emph{PoS} {\bf LATTICE2019}
  (2019) 273}, [\href{https://arxiv.org/abs/1912.04827}{{\tt 1912.04827}}].

\bibitem{Kuramashi:2020meg}
Y.~Kuramashi, Y.~Nakamura, H.~Ohno and S.~Takeda, \emph{{Nature of the phase
  transition for finite temperature $N_{\rm f}=3$ QCD with nonperturbatively
  O($a$) improved Wilson fermions at $N_{\rm t}=12$}},
  \href{http://dx.doi.org/10.1103/PhysRevD.101.054509}{\emph{Phys. Rev.} {\bf
  D101} (2020) 054509}, [\href{https://arxiv.org/abs/2001.04398}{{\tt
  2001.04398}}].

\bibitem{Flores:1983tk}
R.~A. Flores and M.~Sher, \emph{{Is {Coleman-Weinberg} Symmetry Breaking in the
  Standard Model Ruled Out?}},
  \href{http://dx.doi.org/10.1016/0550-3213(84)90344-4}{\emph{Nucl. Phys.} {\bf
  B238} (1984) 702--715}.

\bibitem{Hambye:2018qjv}
T.~Hambye, A.~Strumia and D.~Teresi, \emph{{Super-cool Dark Matter}},
  \href{http://dx.doi.org/10.1007/JHEP08(2018)188}{\emph{JHEP} {\bf 08} (2018)
  188}, [\href{https://arxiv.org/abs/1805.01473}{{\tt 1805.01473}}].

\bibitem{ZinnJustin:1996cy}
J.~Zinn-Justin, \emph{{Quantum field theory and critical phenomena}},
  {\emph{Int. Ser. Monogr. Phys.} {\bf 92} (1996) 1--1008}.

\bibitem{Pisarski:1980ix}
R.~D. Pisarski and D.~L. Stein, \emph{{Critical Behavior of Linear $\phi^4$
  Models With $G \times G^\prime$ Symmetry}},
  \href{http://dx.doi.org/10.1103/PhysRevB.23.3549}{\emph{Phys. Rev.} {\bf B23}
  (1981) 3549--3552}.

\bibitem{Calabrese:2004uk}
P.~Calabrese and P.~Parruccini, \emph{{Five loop epsilon expansion for U(n) x
  U(m) models: Finite temperature phase transition in light QCD}},
  \href{http://dx.doi.org/10.1088/1126-6708/2004/05/018}{\emph{JHEP} {\bf 05}
  (2004) 018}, [\href{https://arxiv.org/abs/hep-ph/0403140}{{\tt
  hep-ph/0403140}}].

\bibitem{Adzhemyan:2021sug}
L.~T. Adzhemyan, E.~V. Ivanova, M.~V. Kompaniets, A.~Kudlis and A.~I. Sokolov,
  \emph{{Six-loop $\varepsilon$ expansion of three-dimensional
  $\text{U}(n)\times \text{U}(m)$ models}},
  \href{https://arxiv.org/abs/2104.12195}{{\tt 2104.12195}}.

\end{thebibliography}\endgroup

\end{document}